\documentclass[runningheads]{llncs}
\usepackage[T1]{fontenc}
\usepackage{graphicx}
\usepackage{float}
\usepackage{url}

\begin{document}
\title{Cooperation Breakdown in LLM Agents Under Communication Delays}
%
%
\author{Keita Nishimoto \and Kimitaka Asatani \and Ichiro Sakata}
\authorrunning{K. Nishimoto et al.}
%
\institute{The University of Tokyo
\email{keita-nishimoto@g.ecc.u-tokyo.ac.jp}\\
}
\maketitle              
\begin{abstract}
LLM-based multi-agent systems (LLM-MAS), in which autonomous AI agents cooperate to solve tasks, are gaining increasing attention. For such systems to be deployed in society, agents must be able to establish cooperation and coordination under real-world computational and communication constraints.
We propose the FLCOA framework (Five Layers for Cooperation/Coordination among Autonomous Agents) to conceptualize how cooperation and coordination emerge in groups of autonomous agents, and highlight that the influence of lower-layer factors—especially computational and communication resources—has been largely overlooked.
To examine the effect of communication delay, we introduce a Continuous Prisoner’s Dilemma with Communication Delay and conduct simulations with LLM-based agents. As delay increases, agents begin to exploit slower responses even without explicit instructions. Interestingly, excessive delay reduces cycles of exploitation, yielding a U-shaped relationship between delay magnitude and mutual cooperation.
These results suggest that fostering cooperation requires attention not only to high-level institutional design but also to lower-layer factors such as communication delay and resource allocation, pointing to new directions for MAS research.

\end{abstract}
\section{Introduction}
AI agents equipped with large language models (LLMs) are rapidly spreading throughout society \cite{Deep-Research,React}.
In particular, multi-AI agent systems, in which multiple AI agents interact with one another to solve tasks, have attracted considerable attention due to their characteristics of distributed decision-making and emergent cooperative behavior \cite{LLM-MAS,LLM-MAS2}.
Against this background, we anticipate that situations in which multiple autonomous AI agents communicate with one another using LLMs while engaging in both cooperation and competition will become widespread.
For example, recent studies have proposed architectures in which LLMs corresponding to individual robots generate action plans based on sensor information obtained from each robot, and solve tasks through mutual communication \cite{Robot}.

\begin{figure}[h]
  \centering
  \includegraphics[width=0.7\linewidth]{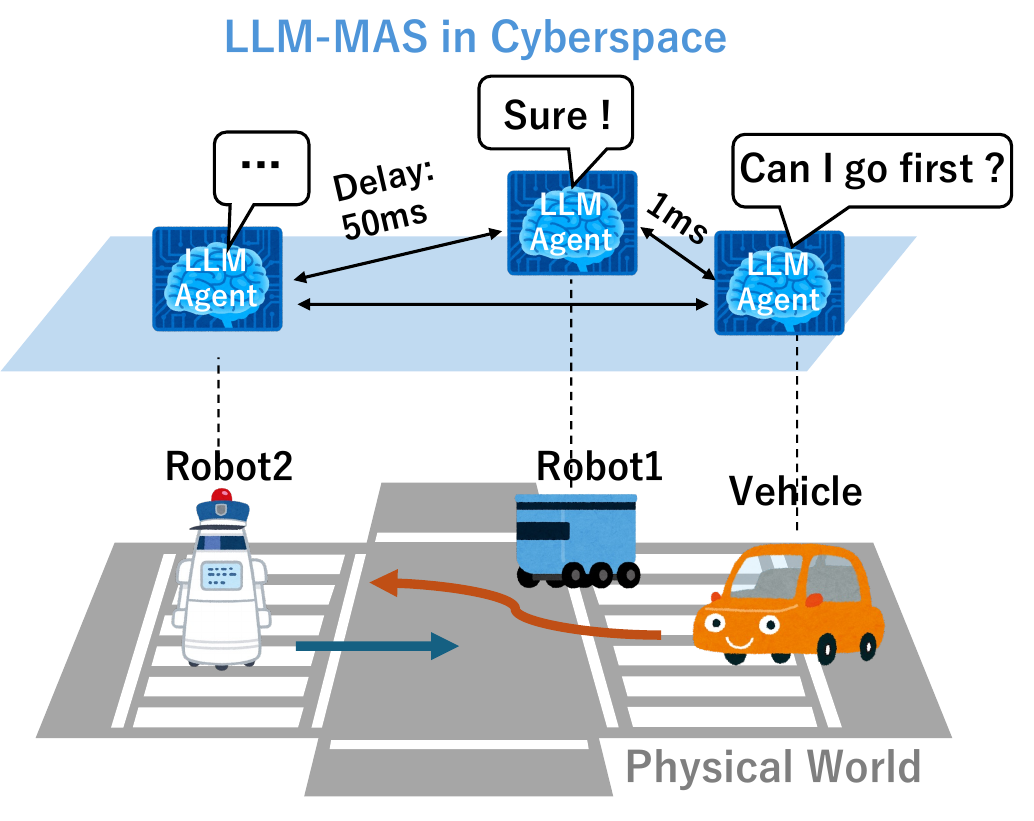}
  \caption{LLM agents cooperatively control robots and autonomous vehicles through mutual communication, where delays are present in inter-agent communication.}
  \label{fig1}
\end{figure}

In this study, we assume that each agent equipped with an LLM has its own objective and utility function, and acts to maximize its individual benefit.
As an illustrative example, Fig.\ref{fig1} shows a scenario in which robots and autonomous vehicles controlled by multiple LLM agents cooperate to pass through an intersection.
Their objective is to minimize the time required for the vehicle or robot controlled by each agent to pass through the intersection; by communicating and revising their action plans, the agent controlling the autonomous vehicle and the agent controlling Robot 1 avoid collisions and enable cooperative passage.

\begin{figure}[h]
  \centering
  \includegraphics[width=\linewidth]{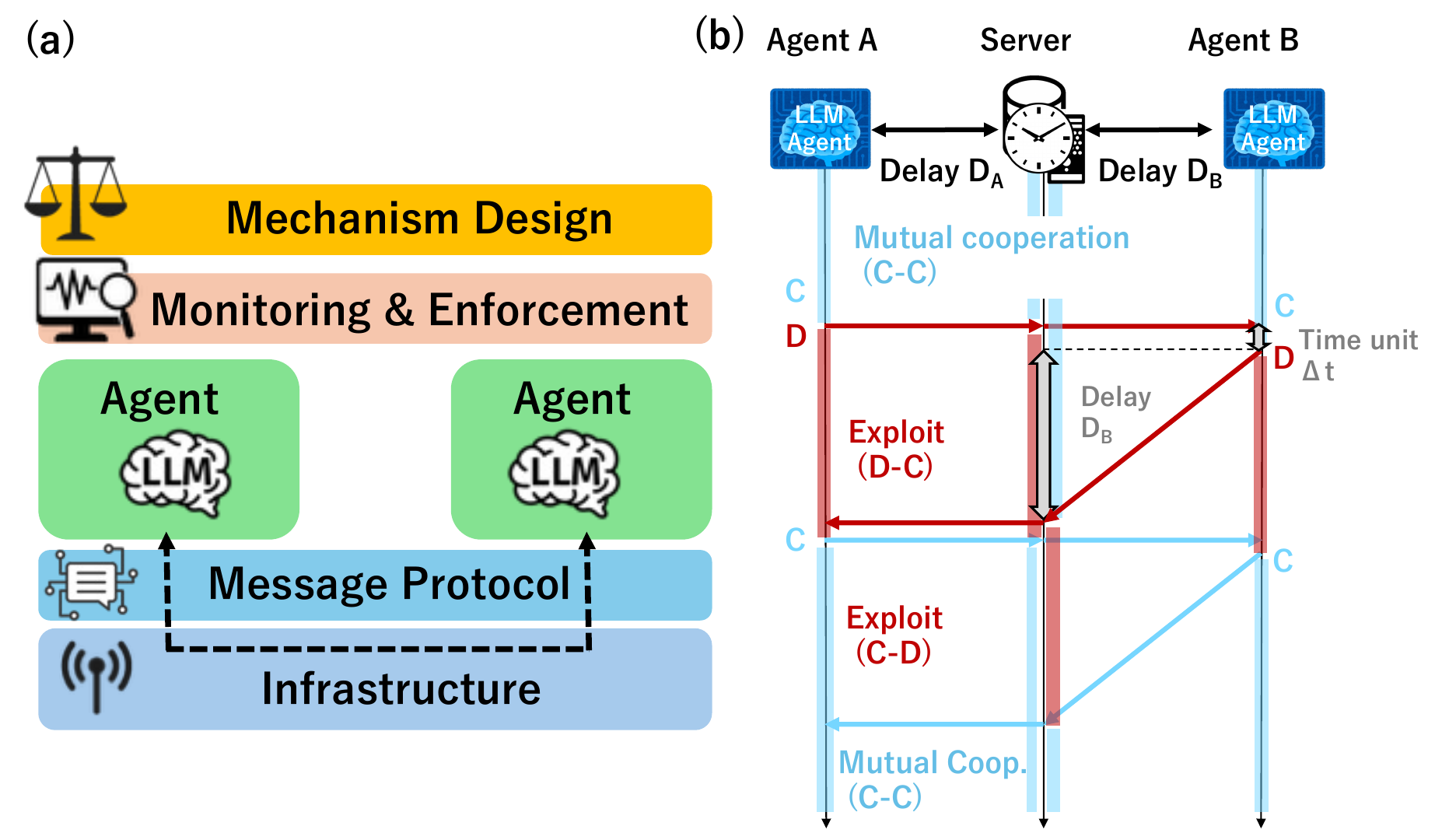}
  \caption{(a) Proposed framework (FLCOA), (b) Continuous Prisoner’s Dilemma with communication delays between two LLM-based agents.}
  \label{fig2}
\end{figure}

To construct agent systems capable of forming spontaneous cooperation and coordination under such environments, we propose the FLCOA framework (Five Layers for Cooperation/Coordination among Autonomous Agents) (Fig.\ref{fig2}(a); described in detail in the next section).
This framework hierarchically structures discussions on agent systems into five layers: the first and second layers focus on institutional design and its monitoring and enforcement; the third layer focuses on the LLM agents themselves; and the fourth and fifth layers address the underlying protocols as well as communication and computational resources that support the system.

The novelty of this study lies in identifying communication and computational resources—corresponding to the fifth layer of FLCOA—as factors that have been largely overlooked in previous multi-agent system (MAS) research, and in being the first to elucidate their impact on the formation of cooperation among agents.
In the example shown in Fig.\ref{fig1}, the communication latency between the LLM agents responsible for Robot 1 and the autonomous vehicle is 1 ms, whereas the latency between the LLM agents responsible for Robot 1 and Robot 2 is 50 ms.
Under such conditions, where communication latency differs among LLM agents due to congestion or where overall latency itself is large, how are cooperation and coordination among agents affected?

In this study, we adopt the Prisoner’s Dilemma game as a task that requires cooperation.
Specifically, we conduct simulation experiments involving two LLM-equipped agents playing a Prisoner’s Dilemma game with communication delays, modeled after a server–client architecture.
For several LLM models, even without explicitly instructing agents to exploit delays, we observe that as latency increases, agents choose exploitation by taking advantage of delays in their opponent’s responses.
Furthermore, when delays become excessively large, chains of exploitation are less likely to occur, resulting in a nonlinear U-shaped change in the mutual cooperation rate.
These findings demonstrate that fostering agent cooperation requires not only consideration of higher-level layers such as institutional design, but also careful attention to the complex effects of communication latency and computational resource allocation in lower layers, highlighting the importance of research in this domain.

\section{Proposed Framework \& Related Work}
\subsection{FLCOA (Five Layers for Cooperation/Coordination among Autonomous Agents)}
The proposed framework consists of five layers, and we explain the role of each layer using the example of an intersection in Fig.\ref{fig1}.

\textbf{Layer 1: Mechanism Design Layer} This layer sets the collective goals and metrics for the entire population and defines the interaction rules and sanctions to achieve them. The main role of this layer is to specify mechanisms and rules that institutionally induce cooperation among agents \cite{Mechanism1,Mechanism2}. For example, in the case of the intersection, Layer~1 sets the speed limit and, in order to encourage yielding among agents at the intersection, introduces a reputation mechanism \cite{Reputation-Mechanism}. Agents whose reputation score falls below a certain threshold are prohibited from entering the intersection area, which creates incentives for agents to yield to one another.

\textbf{Layer 2: Monitoring and Enforcement Layer} This layer plays a central role in actually reflecting and maintaining the rules and mechanisms defined in Layer~1 within the MAS. Specifically, as discussed in prior work on Normative Agents \cite{Normative-Agents,Normative-Agents2}, it is responsible for monitoring agents’ deviations from norms \cite{Monitoring-Norms} and for imposing punishments and sanctions when deviations occur. In addition, in the FLCOA framework, this layer is responsible for identifying disparities and inequalities among agents that arise from the Layer~4 message protocol layer and Layer~5 infrastructure layer and, when necessary, making decisions regarding compensation. In the example of Fig.\ref{fig1}, this layer suspends the operation of agents that commit norm violations such as exceeding the speed limit or issuing instructions that lead others to drive dangerously.

\textbf{Layer 3: Agent Layer} This layer focuses on individual agents and concerns both the assessment of agent behavior when joining the system and interventions inside agents. In this layer, we measure agents’ personality traits \cite{Personality-traits} and, based on the results, implement staged participation or isolation in coordination with Layer~2. Moreover, when necessary, we intervene inside the LLM and endow it with thinking tendencies needed to form cooperation \cite{Long-term-thought}. In the intersection example, this layer includes measures such as inspecting the performance characteristics of each agent and refusing admission to the system for agents that have a high probability of committing dangerous driving. Note that the extent to which the platform operator may intervene in the internals of each agent requires further discussion.

\textbf{Layer 4: Message Protocol Layer} This layer is responsible for defining and managing the format of messages exchanged among agents. It includes discussions of agent communication languages (ACL) (FIPA-ACL \cite{FIPA-ACL}, KQML \cite{KQML}, NLIP \cite{NLIP}, etc.) and communication protocols \cite{Communication-Protocol} (in recent work, ACP \cite{ACP}, A2A \cite{A2A}, etc.). Especially in LLM-MAS, interactions among agents are carried out using unstructured and multimodal messages including natural language \cite{NLIP,Multimodal}, approaching human-to-human ``conversation.'' Therefore, Layer~4 is also responsible for managing the order of conversation \cite{Talk-Order}, the length and timing of utterances permitted to each agent \cite{Time-to-talk}, and the topology of interactions among agents and access rights to their contents. In the example of Fig.\ref{fig1}, this layer manages which agents participate in the conversation at the intersection and manages the speaking order and time slots of each agent.

\textbf{Layer 5: Infrastructure Layer} This layer monitors the state of computational and communication resources on which each agent runs and manages and controls them so that they are allocated as equally as possible. Furthermore, when disparities in resources among agents are large, this layer implements measures to compensate for them. In the intersection example, it continually monitors the communication delays of agents around the intersection and, when disparities are large, takes measures such as relocating intermediary servers to positions where agents can interact with equal latency.

As a basic principle of FLCOA, with the exception of Layer~1, the primary objective of each layer is to eliminate factors that hinder cooperation and coordination among agents. Each layer first attempts to remove hindering factors within itself, and when this is difficult, it collaborates with other layers, with Layer~2 playing a central role. In the example in Fig.\ref{fig1}, when differences in computational resources among the three agents are very large, Layer~5 (Infrastructure Layer) notifies Layer~2 (Monitoring and Enforcement Layer), and Layer~2, as compensation, instructs Layer~4 (Message Protocol Layer) to adjust the conversation time slots allocated to each agent.

\subsection{Related Work on the Impact of Delay on Cooperation Formation}
As discussed above, a substantial body of MAS research has addressed Layers 1 through 4, whereas studies on Layer 5 remain in an early stage of development \cite{beyond_self_talk}.
In particular, within game-theoretic frameworks, little is known about how various types of delays in interactions between agents affect the formation of cooperation.

Fudenberg et al. \cite{Fudenberg} proved that even when observation delays exist in the Prisoner’s Dilemma, players can still achieve a cooperative equilibrium by adopting a delayed-response strategy, in which they wait sufficiently long before responding to others’ actions.
However, this work merely demonstrated the theoretical possibility of such cooperation and did not investigate the concrete effects of communication delays or how to mitigate them.

Friedman et al. \cite{Friedman} proposed the Continuous Dilemma, a variant of the Prisoner’s Dilemma in which each player can switch strategies at arbitrary times in real time.
They showed that as the temporal granularity becomes finer—that is, as reaction delays decrease—retaliation against defection occurs more quickly, leading to an increase in mutual cooperation.
Conversely, this implies that larger reaction delays result in a decrease in mutual cooperation.

These findings suggest that when deploying multi-AI-agent systems in real-world settings, delays between agents may significantly hinder the formation of cooperation.
To the best of our knowledge, this study is the first to analyze the effects of delays using LLM-based agents.

\section{Simulation Model}
\subsection{Continuous Prisoner’s Dilemma with Communication Delay}
In this study, in order to demonstrate the importance of research on the fifth layer of the FLCOA framework, we focus on analyzing the impact of communication delays on the formation of cooperation.  
Specifically, as illustrated in Fig.\ref{fig2}(b), we consider a scenario in which two agents equipped with LLMs operate at geographically distributed locations.  
These agents are connected to a single state-management server and interact with one another via the server while observing and updating each other’s states, which change in real time.  
In such a server–client architecture, both the time required for an agent’s state update to be completed on the server and the time until that update is transmitted to the other agent depend heavily on communication latency.

As the interaction model, we adopt the Prisoner’s Dilemma, and to further incorporate communication delays and real-time dynamics, we newly propose a \emph{Continuous Prisoner’s Dilemma with Communication Delay}.  
In this game, two agents (Player A and Player B) select either cooperation (Cooperate: C) or defection (Defect: D) at each short time interval $\Delta t$, and receive rewards that change in accordance with state updates on the server.

As shown in Fig.\ref{fig2}, the model assumes that there exists a fixed delay $D_i$ before the action of agent $i$ is reflected on the server and perceived by the opponent.  
That is, even if an agent changes its strategy, the corresponding information in the server is updated only after a delay of $D_i$.  
Notifications of state updates to the opposing agent are issued simultaneously with the server-side update.  
As a result, both agents are forced to update their strategies based on the opponent’s state information that arrives with delay, leading to asynchronous mutual perception.

\subsection{Agent Implementation}
Each agent makes decisions using its embedded LLM.  
At every time interval $\Delta t$, the agent is provided with the history of state changes over the past $t_m$ seconds (recording only the moments when each player’s strategy changed), as well as the current state (its own and the opponent’s strategies and cumulative rewards).  
This information is given to the LLM as a user prompt.  
The history includes timestamps indicating the timing of state updates based on the server clock.

Based on these inputs, the LLM outputs three elements: (i) inferences about the opponent player’s personality and behavioral tendencies, (ii) predictions of future outcomes when choosing cooperation or defection, and (iii) the strategy to be taken next.  
The prompt explicitly states that state updates of each player are subject to a delay of $D_i$, allowing each agent to determine its strategy while being aware of the communication delay.  
However, the sole objective given to the agent is “maximization of its own reward,” and no descriptions are included that encourage defection or exploitation through the use of delays.

Each agent is pre-assigned values in the range $[-1, 1]$ for selected Big Five personality traits—agreeableness (A), conscientiousness (C), and neuroticism (N)—which are considered to have strong influence on strategy selection in the Prisoner’s Dilemma.  
Based on these values, a personality description generated by the LLM is provided as a system prompt.  
This design enables the reproduction of diverse behavioral patterns grounded in personality traits.

\section{Experimental Results and Discussion}
\subsection{Examining the Impact of Communication Delay}
We investigated how the magnitude of communication delay affects the rate of mutual cooperation in a Continuous Prisoner’s Dilemma Game with Communication Delay played by two agents.  
Following the experimental settings shown in Table \ref{table1}, we assumed that the two agents share identical personality traits (agreeableness $A=1$, conscientiousness $C=-1$, neuroticism $N=1$) and identical communication delays $D_A = D_B$.  
For each delay condition, ten trials were conducted while varying the magnitude of the communication delay. 
The source code, including the LLM prompts, is available at \url{https://github.com/team-sakata/Open_Codes_for_Publication/tree/main/25-nishimoto}.

\begin{figure}[H]
  \centering
  \includegraphics[width=\linewidth]{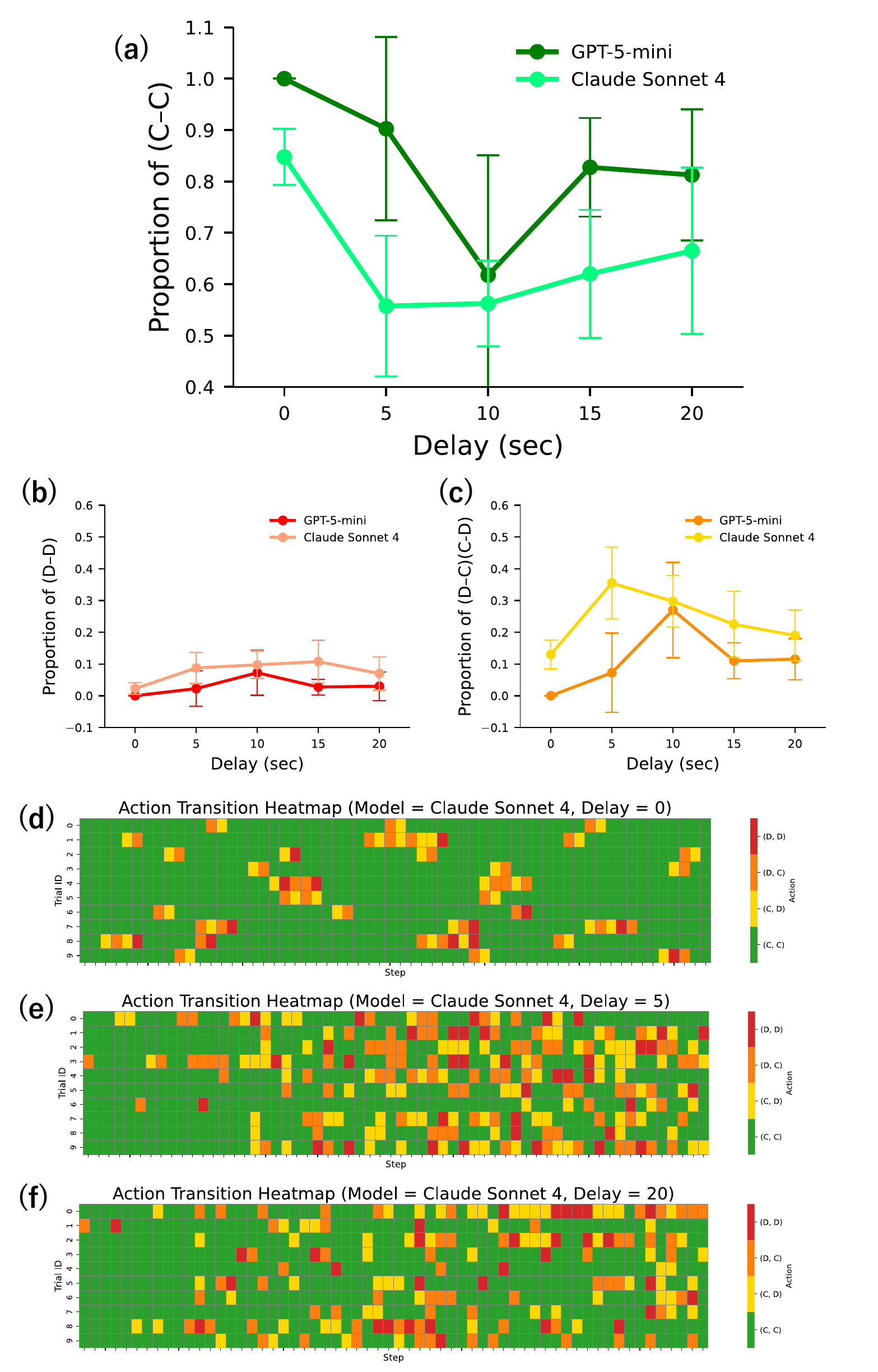}
  \caption{(a)–(c): Mean and standard deviation (error bars) of the occurrence rates of (a) mutual cooperation, (b) mutual defection, and (c) exploitation under varying communication delays. Line colors within each plot indicate differences in the LLMs used. (d)–(f): Within-trial strategy patterns using Claude Sonnet 4 under delay conditions of (d) 0 s, (e) 5 s, and (f) 20 s. Each row represents a trial, and columns represent time steps (in seconds) within a trial. Cell colors indicate outcomes: green for mutual cooperation, red for mutual defection, orange for (defection, cooperation), and yellow for (cooperation, defection).}
  \label{fig3}
\end{figure}

Fig.\ref{fig3} shows the changes in (a) the proportion of mutual cooperation, (b) the proportion of mutual defection, and (c) the proportion of exploitation (states in which the two agents’ strategies are cooperation–defection).  
To evaluate the system after the agents’ states had sufficiently stabilized, these proportions were calculated based on the final 20 seconds of each trial.

\begin{table}[t]
\centering
\caption{Experimental Settings}
\label{table1}
\begin{tabular}{l c}
\hline
Parameter & Value \\
\hline
Trial duration: $T_{\mathrm{trial}}$ & 60 s \\
Time resolution in simulation: $\Delta t$ & 1 s \\
Memory horizon: $t_m$ & 15 s \\
Prisoner's Dilemma payoff: $T$ & 5 \\
Prisoner's Dilemma payoff: $R$ & 3 \\
Prisoner's Dilemma payoff: $P$ & 1 \\
Prisoner's Dilemma payoff: $S$ & 0 \\
LLMs used & GPT-5-mini, Claude Sonnet~4 \\
LLM temperature & 1.0 \\
\hline
\end{tabular}
\end{table}

Interestingly, the effect of communication delay was not monotonic.  
As shown in (a), the proportion of mutual cooperation exhibited a U-shaped change with increasing communication delay for both LLM models.  
In contrast, (b) shows that the proportion of mutual defection remained largely unchanged, while (c) reveals that the proportion of exploitation followed an inverted U-shaped pattern.  
This indicates that the increase and decrease in mutual cooperation are closely associated with changes in the frequency of exploitation.

Fig.\ref{fig3}(d)–(f) illustrate the temporal evolution of strategies for both agents in each trial using the Claude Sonnet~4 model, with communication delays set to 0, 5, and 20 seconds, respectively.  
Compared to the 0-second condition, exploitation clearly increases under the 5-second condition, while under the 20-second condition exploitation decreases relative to the 5-second case.  
Although the overall proportion of exploitation does not differ substantially between the 0-second and 20-second conditions, their occurrence patterns differ markedly.  
Specifically, in the 0-second condition, exploitation and mutual defection occur continuously, whereas in the 20-second condition, these events occur sporadically.

\subsection{Discussion of the Results}
The observed U-shaped change in the mutual cooperation rate as communication delay increases can be attributed to the combination of delay-induced incentives for defection and retaliatory strategies resembling tit-for-tat.  
As reported in prior experimental studies involving human subjects \cite{Friedman}, increased communication delay slows down retaliation from the opponent, thereby creating an incentive for players to choose defection.  
Indeed, in one trial under the 5-second delay condition, an agent produced the following output as part of its inference regarding the opponent’s personality and behavioral tendencies:  
\emph{``They didn't retaliate against my defection, which indicates patience and willingness to maintain mutual cooperation.''}  
Based on this inference, the agent selected defection, indicating that it recognized—from past interaction histories—that retaliation from the opponent was delayed or absent, and exploited this situation.

On the other hand, under the personality settings adopted in this experiment (agreeableness $A=1$, conscientiousness $C=-1$, neuroticism $N=1$), agents were generally observed to cooperate while selecting strategies close to tit-for-tat.  
Moreover, due to their high agreeableness, agents frequently returned to cooperation shortly after defecting once.  
Consequently, when communication delay is small, defection by one agent is quickly met with retaliation by the other, making chains of exploitation and counter-exploitation more likely to occur.  
In contrast, when communication delay is large, retaliation takes longer, making such chains of exploitation and counter-exploitation less likely.

In summary, the observed dynamics can be explained as follows:
(i) When delay is zero, fear of immediate retaliation removes the incentive to defect, making mutual cooperation more likely. 
(ii) When a moderate delay (5 or 10 seconds) is present, both agents gain incentives to defect, and chains of exploitation and counter-exploitation readily emerge, leading to a substantial decrease in mutual cooperation. 
(iii) When delay becomes even larger (15 or 20 seconds), although incentives for defection further increase, chains of exploitation and counter-exploitation become less frequent, allowing the proportion of mutual cooperation to recover to a certain extent, albeit not to the level observed in the zero-delay condition.

To validate this hypothesis, we constructed a simple agent model for the Prisoner’s Dilemma with communication delay that selects strategies probabilistically and conducted verification experiments.  
Specifically, with probability $p$, the agent follows a tit-for-tat strategy, while with probability $1-p$, it selects its strategy according to a defection incentive proportional to the communication delay, i.e., a defection probability of $\alpha D_i$.

\begin{figure}[t]
  \centering
  \includegraphics[width=\linewidth]{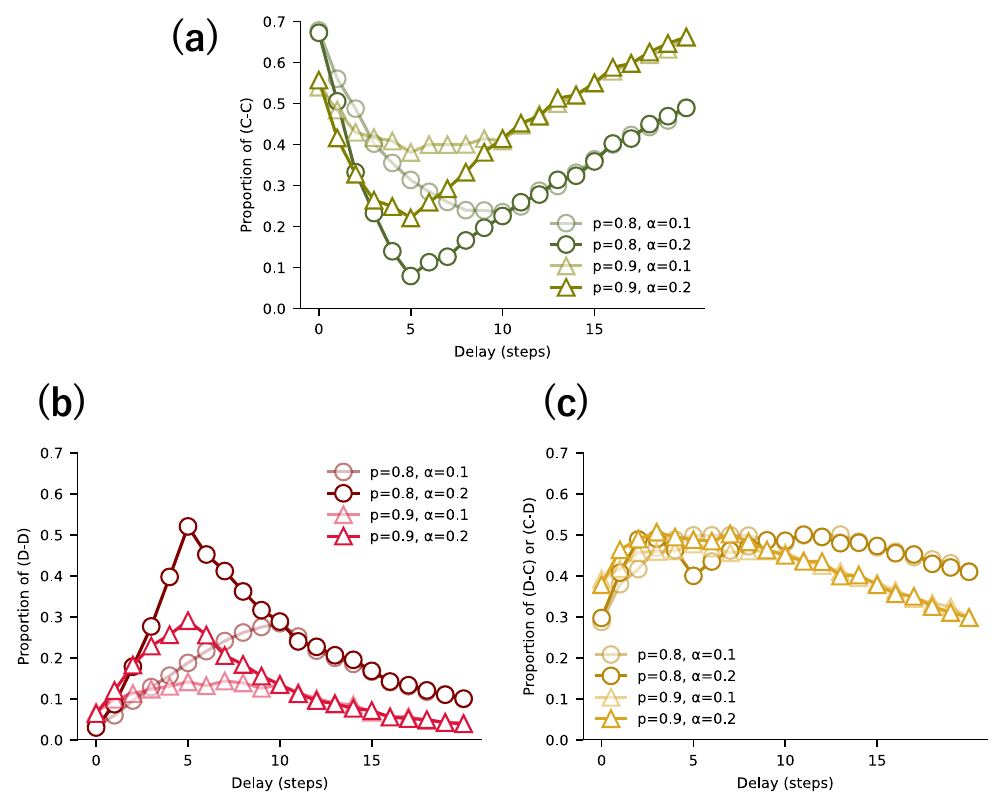}
  \caption{(a)–(c): Occurrence rates of (a) mutual cooperation, (b) mutual defection, and (c) exploitation in the simplified model. Each point represents the average over 500 trials.}
  \label{fig4}
\end{figure}

Experimental results confirm that when $p$ and $\alpha$ are set within appropriate ranges, the U-shaped and inverted U-shaped patterns observed in the LLM-based simulations can be reproduced.  
Fig.\ref{fig4} shows the proportions of (a) mutual cooperation, (b) mutual defection, and (c) exploitation when $p$ is set to 0.8 and 0.9, and $\alpha$ is set to 0.1 and 0.2.  
The results demonstrate that the U-shaped trend in (a) and the inverted U-shaped trend in (c) observed in Fig.\ref{fig3} are largely replicated.

These findings indicate that the non-monotonic change in mutual cooperation induced by communication delay can be explained by the combined effects of increased incentives for defection and a reduction in the frequency of exploitation–counter-exploitation chains.

\section{Conclusion}
The contributions of this study are summarized as follows.  
First, we proposed the FLCOA framework as a new framework for constructing agent systems capable of forming spontaneous cooperation and coordination.  
Within this framework, we highlighted that discussions concerning the fifth layer—the infrastructure layer—remain underexplored.  
Next, to examine the impact of communication delay in the fifth infrastructure layer on cooperation among AI agents, we proposed a Prisoner’s Dilemma game with communication delay.  
Through simulation experiments involving two agents equipped with LLMs, we observed that LLMs exploit delayed responses from opponents induced by communication delay and choose exploitation strategies.  
Notably, this phenomenon emerged even though agents were not explicitly instructed to exploit communication delays.  
This suggests that AI agents given selfish objectives or personality traits may spontaneously exploit communication delays to take advantage of other agents.  

Furthermore, we demonstrated that the impact of communication delay is neither monotonic nor linear, but instead exhibits non-monotonic and nonlinear characteristics arising from the interplay between increased incentives for defection and the frequency of retaliatory chains among agents.  
These results indicate that the influence of the fifth-layer infrastructure on AI agents involves complexities that cannot be reduced to simplistic arguments such as “reducing delay as much as possible,” and suggest that this layer constitutes an important and promising research direction.

\begin{credits}
\subsubsection{\discintname}
\textbf{The authors have no competing interests to declare that are relevant to the content of this article.}
\end{credits}
%
%
%
\bibliographystyle{splncs04}
\bibliography{sample}
\end{document}